\documentclass{icrc2009}

\usepackage{graphicx}   
\usepackage[caption=false]{caption}    
\usepackage[font=footnotesize]{subfig} 
\usepackage{fixltx2e}
\usepackage{url}

\newcommand{\shorttitle}[1]%
{\markboth{Proceedings of the 31\MakeLowercase{$^{st}$} ICRC, {\L}\'{o}d\'{z} 2009}{#1} }
\newcommand{\etal}{\MakeLowercase{\textit{et al.}}} 

\begin{document}
\title{Gamma ray astronomy with \textsc{Antares}}

\author{\IEEEauthorblockN{Goulven Guillard\IEEEauthorrefmark{1}, for the \textsc{Antares} Collaboration}\\
\IEEEauthorblockA{\IEEEauthorrefmark{1}Institut Pluridisciplinaire Hubert Curien (CNRS/IN2P3)\\\& Universit\'e Louis Pasteur, Strasbourg, France}}

\shorttitle{Goulven Guillard - Gamma ray astronomy with \textsc{Antares}}
\maketitle

\begin{abstract}
It has been suggested that underwater neutrino telescopes could detect muons from gamma ray showers. \textsc{Antares}' ability to detect high energy muons produced by TeV photons is discussed in the light of a full \textsl{Monte Carlo} study. It is shown that currently known sources would be hardly detectable.
\end{abstract}

 
\section{Introduction}
The last decade has been fruitful in terms of high energy astronomy. More than 80 gamma rays sources have been found to emit in the TeV range~\cite{catalog}, thanks to Imaging Atmospheric \v{C}erenkov Telescopes (IACTs) such as \textsc{Cangaroo}, \textsc{HEGRA}, H.E.S.S, \textsc{Magic} or \textsc{Veritas}/Whipple~\cite{IACTs}.

There have also been many discussions about the possibility to detect muons produced by high energy gamma rays in underground, underice or underwater neutrino telescopes~\cite{gammarays}\cite{gammaraysok}\cite{Kudryavtsev}. In contrast to upward-going muons from neutrinos, which are the primary purpose of such a telescope, downward-going muons from gamma rays suffer from a high atmospheric muon background. Therefore the sensitivity of a neutrino telescope to gamma ray induced muons is quite lower than IACTs'. However, it has the advantage of monitoring continuously all directions. In addition to their physics potential, muons from gamma rays may also offer calibration benefits in terms of pointing accuracy and angular resolution.

Gamma ray showers are believed to be muon poor, but there are at least three processes by which a photon can produce muons\,: photoproduction, muon pair production and charm decay. The first process involves the (semi)leptonic decay of a pion or a kaon produced by the interaction of the photon with an atmospheric nucleus. Such muons are said to be conventional. The second process is self-explanatory, and its final particles are referred to as direct muons. The final case corresponds to the (semi)leptonic decay of a photoproduced charm meson, and secondary muons are called prompt muons. The prompt muon production was not taken into account in this work since it was not implemented in the software used for the \textsl{Monte Carlo} production. The charm production involves QCD processes that are not fully understood, but measurements at HERA have shown that at photon energies of several TeV charm production is significant~\cite{charm}.

Some calculations have estimated that the muon flux from gamma ray sources could be sufficient for neutrino telescopes to detect them. However, most attempts to estimate this muon flux rely on one-dimensional analytic models, and do not take into account the muon propagation from sea level to the detector and the detector sensitivity. A first attempt to estimate the underwater flux using a \textsc{Monte Carlo} simulation, without considering detection efficiency, has found gamma ray sources to be hardly detectable by a neutrino telescope~\cite{Kudryavtsev}.

In this paper, a full \textsl{Monte Carlo} simulation, including \v{C}erenkov light detection in realistic background conditions and track reconstruction, is presented, within the \textsc{Antares} framework. The expected number of events from the main sources of interest are presented.

\section{The \textsc{Antares} detector}

The Mediterranean sea currently houses the first operational undersea neutrino telescope, and also the largest neutrino telescope in the Northern hemisphere, namely \textsc{Antares} (Astronomy with a Neutrino Telescope and Abyss environmental RESearch)~\cite{ANTARES}. Its full configuration has been completed in May 2008, though data has been taken with partial detector configurations since the first line was in water, in March 2006.

\textsc{Antares}' main focus is to detect astrophysical neutrinos, thanks to the \v{C}erenkov light produced in water by muons resulting from the interactions of neutrinos with the Earth. Because of the atmospheric muon background, \textsc{Antares} field of view is the Southern hemisphere, and in particular the Galactic center.

Installed at 40\,km off Toulon, in France (40$^{\circ}$50$^{\prime}$\,N, 6$^{\circ}$10$^{\prime}$\,E), \textsc{Antares} comprises 12 vertical detection lines positioned on the Mediterranean sea bed, at about 2500\,m depth. Each line hosts up to 25 floors of three 10$^{\prime\prime}$ photomultiplier tubes (PMTs) regularly distributed over 350\,m, the lowest floor being 100\,m above the sea bed. On a given floor, the three PMTs are placed symmetrically around the vertical string and look downwards at 45$^{\circ}$ in order to optimize the collection of \v{C}erenkov photons from upgoing muons rather than from downgoing muons~\cite{OMs}.

The lines are separated from each other by approximately 70\,m, and set as a regular octagon surrounding a square. An instrumented line intended to monitor the environmental conditions completes the apparatus.

The sea current induced displacements of the lines with regard to their vertical axis do not exceed a few meters, and are monitored in real time using compasses, tiltmeters and hydrophones hosted on each line. A position accuracy of about 10\,cm for each PMT is obtained.

An electro-optical cable transfers the electronic readout of the whole detector to shore, where digitized informations are processed in a computer farm.

\section{Monte Carlo simulation}

Extensive Air Shower have been simulated using Corsika v6.720~\cite{CORSIKA}. High energy hadronic interactions are modeled through QGSJET01~\cite{hadronic}, while electromagnetic interactions are processed through EGS4~\cite{EGS4}. QGSJET01 is found to be the most conservative model in comparison to SIBYLL, VENUS and QGSJET-II~\cite{hadronic}, regarding the number of photons creating high energy muons (using VENUS leads to a 7\% rise, assuming a E$_\gamma^{-1}$ flux, in the [1;100]\,TeV range). However, the effect of this increase at the depth of \textsc{Antares} still has to be investigated, the muon range being energy dependent. The energy range considered in the present work goes from 1\,TeV to 100\,TeV.

MUSIC has been used for the propagation of muons in water~\cite{MUSIC}. The \textsc{Antares} \textsl{Monte Carlo} simulation chain then allows for simulation of \v{C}erenkov light in the detector, taking into account, in particular, the water properties and the PMTs angular acceptance~\cite{MC}. It also allows for the addition of realistic bioluminescence background using real data streams.

In this work the data used for the bioluminescence background corresponds to golden running conditions\,: runs are selected where the baseline of raw counting rates and the fraction of bursts\footnote{The burst fraction is defined as the ratio between the time when the counting rate is higher than 250\,kHz and the overall time.} are low (about 60\,kHz and less than 20\%, respectively).

Finally, the events are reconstructed using \textsc{Antares} standard reconstruction strategy. It has to be noticed that this strategy is optimized for upgoing events. The results presented here might thus be slightly enhanced using a dedicated strategy. On the other hand, the cut made on the reconstruction quality is very loose, so the effect of hardening the quality cut may compensate the effect of improving the reconstruction strategy.

\section{Sources of interest}

In order to have a reasonable probability to reach the depth of the \textsc{Antares} detector, a downgoing vertical muon must be more energetic than 1\,TeV at sea level\,: the muon probability to survive to a 2200\,m depth in water is less than 70\% for a 1\,TeV muon (13\% at 700\,GeV). Hence only TeV gamma ray sources may be seen by \textsc{Antares}. More than 80 gamma ray sources have been detected in the TeV range by IACTs~\cite{catalog}. However, not all of them are good candidates for \textsc{Antares}\,: most of them are located in the galactic plane, which is not in \textsc{Antares} field of view\footnote{The gamma rays field of view being the Northern hemisphere, as opposed to the neutrinos field of view.}. Moreover, weak and/or soft fluxes are not likely to produce enough muons.

Fortunately, several of the most powerful sources are visible by \textsc{Antares}, including the so-called ``standard candle'', the Crab pulsar. Characteristics of the most interesting candidates in terms of fluxes and visibility are summarized in table~\ref{tab:sources}. In addition to the Crab, three extragalactic sources have been selected.

Though most of these sources are variable or flaring sources, they are known to have long periods of high activity, which make them more promising over a long period than most steady sources~\cite{1ES}\cite{Mkn421}\cite{Mkn501}.

\begin{table}[!h]
\centering
\begin{tabular}{lccc}\hline
source		& visibility	& mean zenith	& type	\\\hline\hline
Crab		&	62\%	&	51.7	& PWN	\\\hline
1ES 1959+650	&	100\%	&	49.7	& HBL	\\\hline
Mkn 501		&	78\%	&	49.4	& HBL	\\\hline
Mkn 421		&	76\%	&	49.2	& HBL	\\\hline
\end{tabular}
\caption{\textsl{Characteristics of \textsc{Antares} best gamma ray sources. HBL stands for High frequency peaked BL Lac object, and PWN for Pulsar Wind Nebula.}}
\label{tab:sources}
\end{table}

\section{Simulation results}

\begin{table*}[!th]
\centering
\begin{tabular}{lcccccc}\hline
source		& $\mathrm{F_{\gamma}^{atm}}$ 	& $\mathrm{F_{\gamma}^{sea}(\times10^{-3})}$	& $\mathrm{N^{det}_{5{\scriptscriptstyle +}}}$& $\mathrm{N_{10{\scriptscriptstyle +}}^{det}}$&  $\mathrm{N^{reco}}$\\[0.1cm]\hline\hline
Crab		&	5-8	&	0.2-0.8		&	30-70		&	20-45		&	1-4	\\
		&		&	\emph{0.4-1}	&	\emph{20-50}	&	\emph{15-35}	&	\emph{1-3}	\\\hline
1ES 1959+650	&	0.8-30	&	0.1-2		&	3-150		&	2-100		&	0.2-8	\\
		&		&	\emph{0.1-2}	&	\emph{2-100}	&	\emph{1-70}	&	\emph{0.1-5}	\\\hline
Mkn 421		&	1.5-45	&	0.1-4		&	5-330		&	3-230		&	0.2-20	\\
		&		&	\emph{0.1-5}	&	\emph{2-260}	&	\emph{2-190}	&	\emph{0.1-15}	\\\hline
Mkn 501		&	1.5-40	&	0.1-15		&	6-1350		&	4-950		&	0.3-90	\\
		&		&	\emph{0.1-15}	&	\emph{3-1200}	&	\emph{2-880}	&	\emph{0.1-80}	\\\hline
\end{tabular}
\caption{\textsl{Number of photons/muons produced by several gamma ray sources at different levels, during one year, assuming a 100\% visibility, for primaries in the [1;100]\,TeV energy range. When relevant, straight font corresponds to a 10 degrees zenith angle, while italic stands for a 40 degrees zenith angle. Fluxes of photons at the top of the atmosphere ($\mathrm{F_{\gamma}^{atm}}$) and at sea level ($\mathrm{F_{\gamma}^{sea}}$) are expressed in m$^{-2}$. $\mathrm{N^{det}_{X{\scriptscriptstyle +}}}$ is the number of photons which produce more than $X$ hits on the detector PMTs, and $\mathrm{N^{reco}}$ corresponds to the number of reconstructed events in realistic bioluminescence conditions.}}
\label{tab:results}
\end{table*}

The number of detected and reconstructed events depends on several parameters, such as the precise level of background, the trigger strategy, the reconstruction strategy and the source flux parametrization. Therefore only range estimates are given. They are reported in table~\ref{tab:results}, assuming a 100\% visibility over one year. Most excentric parametrizations have been omitted.

It is found that only a few photons can be expected to be seen by \textsc{Antares}\,: less than ten events per year are reconstructed for the Crab in realistic conditions, though a few tens produce hits on the detector.

In comparison, a rough estimate on data with similar bioluminescence conditions gives $1.1\times10^5$ (resp. $4.1\times10^4$) reconstructed background events (atmospheric muons) within a one degree cone of the 10 degrees zenith angle direction (resp. 40 degrees), and $2.1\times10^6$ (resp. $7.3\times10^5$) background events within a 5 degrees cone.

It seems thus not reasonable to expect \textsc{Antares} to extract any gamma signal from the background under these conditions for any known flux. Though Markarian 501 may seem promising, the upper limit actually corresponds to high state fluxes parametrizations on dayscale variations~\cite{Mkn501}. A more precise selection of the fluxes and a study of the significance of the expected number of muons are still to be done.

However, these estimates are not so bad as they seem. First, the simulation is conservative in terms of photoproduction cross-section~\cite{sigmaph} and does not take muon production from charm decay into account. In addition, background discrimination has not yet been investigated. In particular, the muon poorness of gamma ray showers may help to reduce the atmospheric background\,: by rejecting multimuon events, one can improve both the signal to noise ratio and the angular resolution. If achievable, the muon pair tagging may also improve the background rejection. Moreover, a dedicated reconstruction strategy could increase the number of detected photons. Finally, galactic sources such as the Crab are not subject to the universe opacity above 100\,TeV, and the extension of their spectra to higher energy could lead to reasonable numbers of detectable photons. Short and powerful bursts are not to be excluded either, the associated background being in such cases almost negligible.

\section{Conclusion}

A complete \textsl{Monte Carlo} study has been processed in order to estimate \textsc{Antares}' ability to detect downgoing muons from gamma ray sources.

It has been found that \textsc{Antares} is not likely to detect any of the currently known sources, unless they show some unexpected behaviour. However the conservative estimates computed in this work show that the gamma ray astronomy field is not completely out of reach of underwater neutrino telescopes, at least for the next generation of detectors.

This study will be refined and extended to the km$^3$-scale successor of \textsc{Antares}, namely KM3NeT, which is currently being designed~\cite{KM3NET}.

\end{document}